\documentclass[a4paper, 12pt, oneside]{article}
\usepackage[cp1251]{inputenc}
\usepackage[english, russian]{babel}
\usepackage{ graphics,amsfonts, amsmath,bm,amssymb, array}
\usepackage[dvips]{graphicx}
\usepackage[flushleft,small]{caption2}
\makeatletter
 \renewcommand{\@biblabel}[1]{#1.\hfill}

\renewcommand{\Re}{\mathop{\rm Re\,}}

\newcommand{\Kn}{\mathop{\rm Kn\,}}
\makeatother {

\begin{document}
\thispagestyle{empty} \large
\renewcommand{\abstractname}{\, }
\renewcommand{\refname}{\begin{center}\normalsize \rm
СПИСОК ЛИТЕРАТУРЫ\end{center}}

\begin{center}
 {\bf Analytical solution of second Stokes problem of behaviour of rarefied gas with Cercignani boundary accomodation conditions
}\\ \bigskip

\begin{center}
\bf A. V. Latyshev and A. A. Yushkanov
\end{center}

\bigskip
{\it 105005 Moscow, ul. Radio 10а,\\
 Moscow State Regional University\\ E-mail: avlatyshev@mail.ru;
yushkanov@inbox.ru }\\
\end{center}\bigskip

The second Stokes problem about behaviour of rarefied gas filling half-space is analytically solved. A plane,
limiting half-space, makes harmonious fluctuations
in the plane. The kinetic BGK--equation (Bhat\-nagar,
Gross, Krook) is used. The boundary accomodation conditions of Cercignani of reflexion gaseous molecules from a wall are considered. Distribution function of the gaseous molecules is constructed. The velocity of gas in half-space is found, also its
value direct at a wall is found. The force resistance operating from gas on border is found. Besides, the capacity of dissipation of the energy falling to unit of area of the fluctuating plate limiting gas is obtained.


Аналитически решена вторая задача Стокса о поведении разреженного газа, заполняющего полупространство. Плоскость,
ограничивающая полупространство, совершает гармонические колебания
в своей плоскости. Используется кинетическое БГК--уравнение (Бхатнагар,
Гросс, Крук). Рассматриваются граничные аккомодационные условия Черчиньяни отражения
молекул газа от стенки. Построена функция распределения газовых
молекул. Найдена скорость газа в полупространстве, отыскивается
ее значение непосредственно у стенки. Найдена сила сопротивления,
действующая со стороны газа на границу. Кроме того, отыскивается
мощность диссипации энергии, приходящаяся на единицу площади колеблющейся
пластины, ограничивающей газ.

{\bf Ключевые слова:} собственные решения, непрерывный и дискретный спектр,
точное решение, скорость газа, сила трения, диссипация энергии.

\begin{center}
ВВЕДЕНИЕ
\end{center}

Впервые задача о поведении сплошной среды над стенкой, колеблющейся в своей
плоскости, была рассмотрена Дж. Г. Стоксом \cite{Stokes} в середине XIX
столения. Сейчас такую задачу называют второй задачей
Стокса \cite{SK-2007}--\cite{DudkoD}.

Задача о поведении газа над движущейся поверхностью в последние годы
привлекает пристальное внимание \cite{SK-2007}--\cite{DudkoD}. Это связано с
развитием современных технологий \cite{Steinhell}, в частности,
технологий наноразмеров \cite{Karabacak}, \cite{Cleland}.
В \cite{SK-2007}--\cite{DudkoD} эта задача решалась численными или
приближенными методами.

Подробная история этой проблемы изложена в \cite{DudkoD} и \cite{ALY-1}.

В диссертации \cite{DudkoD} были предложены два решения второй задачи Стокса,
учитывающие весь возможный диапазон коэффициента аккомодации
тангенциального импульса. Эти решения отвечают соответственно
гидродинамическому и кинетическому описанию поведения газа над
колеблющейся поверхностью в режиме со скольжением.

В наших работах \cite{ALY-1}--\cite{ALY-3} для второй задачи Стокса
отыскиваются собственные функции и соответствующие собственные значения,
отвечающие как дискретному, так и непрерывному спектрам. Исследована
структура дискретного и непрерывного спектров. Развивается математический
аппарат, необходимый для аналитического решения задачи и приложений.
Дается аналитическое решение рассматриваемой задачи с диффузными граничными
условиями.

В настоящей работе строится аналитическое решение второй задачи Стокса с аккомодационными граничными условиями Черчиньяни \cite{Cerc73}.
На основе аналитического решения вычисляется скорость газа в
полупространстве и непосредственно у колеблющейся границы, найдена сила
трения, действующая со стороны газа на колеблющуюся пластину, а также
находится диссипация энергии пластины.

\begin{center}
1. ПОСТАНОВКА ЗАДАЧИ
\end{center}
Пусть разреженный одноатомный газ занимает полупространство $x>0$
над плоской твердой поверхностью, лежащей в плоскости $x=0$.
Поверхность $(y,z)$ совершает гармонические колебания вдоль оси $y$
по закону $u_s(t)=u_0e^{-i\omega t}$. Требуется построить функцию
распределения газовых молекул $f(t,x,\mathbf{v})$, найти скорость газа
$u_y(t,x)$ и другие макропараметры задачи: силу трения и мощность
диссипации энергии.
Линеаризация задачи проведена при условии, что скорость газа много меньше
тепловой: $|u_y(t,x)|\ll v_T$,
где $v_T=1/\sqrt{\beta} $ -- тепловая скорость молекул $(\beta=m/(2kT))$,
имеющая порядок скорости звука. Здесь $m$ -- масса молекулы газа, $T$ -- его
температура, $k$ -- постоянная Больцмана.

Функция распределения ищется в виде
$f=f_0(1+\varphi e^{-i\omega t})$, где $f_0$ -- абсолютный максвеллиан,
$f_0=n(\beta/\pi)^{3/2}\exp(-\beta v^2)$, $n$ -- концентрация
(числовая плотность) газа.
Рассмотрим линеаризованное кинетическое уравнение
$$
v_x\dfrac{\partial \varphi}{\partial x}+(\nu-i \omega)\varphi(x,\mathbf{v})=
\dfrac{\nu m}{kT}v_yu_y(t,x),
\eqno{(1.1)}
$$
где $u_y(t,x)$ -- скорость газа,
$$
u_y(t,x)=\dfrac{1}{n}\int v_yf(t,x,\mathbf{v})d^3v,
$$
Здесь $\nu=1/\tau$ -- частота столкновений газовых молекул, $\tau$ --
время между двумя последовательными столкновениями молекул.
Концентрация газа и
его температура считаются постоянными в линеаризованной постановке задачи.

Введем безразмерные скорости и параметры: безразмерную скорость молекул:
$\mathbf{C}=\sqrt{\beta}\mathbf{v}$, безразмерную
массовую скорость $U_y(t,x)=\sqrt{\beta}u_y(t,x)$, безразмерные координату
и время $x_1=\nu \sqrt{\beta}x_1$ и $t_1=\nu t$ и
безразмерную скорость колебаний пластины $U_s(t)=U_0e^{-i\omega t}$,
где $U_0=\sqrt{\beta}u_0$ -- безразмерная амплитуда скорости колебаний
границы полупространства. Тогда уравнение (1.1) может быть записано в виде:
$$
C_x\dfrac{\partial \varphi}{\partial x_1}+z_0\varphi(x_1,\mathbf{C})=
{2C_y}U_y(t_1,x_1),
\eqno{(1.2)}
$$
где $z_0=1-i\omega_1, \omega_1=\omega\tau=\omega/\nu$,
$$
U_y(t_1,x_1)=\dfrac{e^{-i\omega_1t_1}}{\pi^{3/2}}\int
\exp(-C^2)C_y\varphi(x_1,\mathbf{C})d^3C.
\eqno{(1.3)}
$$

С помощью (1.3) уравнение (1.2) записывается в виде:
$$
C_x\dfrac{\partial \varphi}{\partial x_1}+z_0\varphi(x_1,\mathbf{C}) =
\dfrac{2C_y}{\pi^{3/2}}
\int\exp(-{C'}^2)C_y'\varphi(x_1,\mathbf{C'})\,d^3C'.
\eqno{(1.4)}
$$

Сформулируем 
аккомодационные граничные условия Черчиньяни \cite{Cerc73}, записанные относительно
функции $\varphi(x_1,\mathbf{C})$:
$$
\varphi(0,\mathbf{C})=2C_y[U_0q+d],\quad C_x>0,
\eqno{(1.5)}
$$
$$
1-q=-\dfrac{\displaystyle2d \int\limits_{C_x>0}f_0C_xC_y^2d^3C}
{\displaystyle\int\limits_{C_x<0}f_0\varphi(0,\mathbf{C})
C_xC_yd^3C},
\eqno{(1.6)}
$$
и
$$
\varphi(x_1\to+\infty,\mathbf{C})=0.
\eqno{(1.7)}
$$
Здесь $q$ -- коэффициент аккомодации, $0 \leqslant q \leqslant 1, d$ --
неизвестная постоянная, подлежащая нахождению из условий задачи.

Итак, граничная задача о колебаниях газа сформулирована полностью и состоит
в решении уравнения (1.4) с граничными условиями (1.5)--(1.7).

Следуя Черчиньяни \cite{Cerc73}, положим далее
$\varphi=C_yh(x_1,\mu), \mu=C_x$. Уравнение (1.4) упрощается при этом и принимает
вид:
$$
\mu\dfrac{\partial h}{\partial x_1}+z_0h(x_1,\mu)=\dfrac{1}{\sqrt{\pi}}
\int\limits_{-\infty}^{\infty}\exp(-{\mu'}^2)h(x_1,\mu')d\mu'.
\eqno{(1.8)}
$$
Граничные условия также упрощаются:
$$
h(0,\mu)=2(U_0q+d), \qquad \mu>0,
\eqno{(1.9)}
$$
$$
\int\limits_{-\infty}^{\infty}e^{-\mu^2}\mu h(0,\mu)d\mu=q\Big(U_0-\dfrac{d}{1-q}\Big).
\eqno{(1.10)}
$$
$$
h(+\infty,\mu)=0.
\eqno{(1.11)}
$$
Далее будем решать задачу (1.8)--(1.11).

\begin{center}
2. СОБСТВЕННЫЕ РЕШЕНИЯ
\end{center}
Разделение переменных в уравнении (1.8) осуществляется следующей подстановкой
$$
h_\eta(x_1,\mu)=\exp\Big(-\dfrac{x_1z_0}{\eta}\Big)\Phi(\eta,\mu),
\eqno{(2.1)}
$$
где $\eta$ -- параметр разделения, или спектральный параметр.

Подставляя (2.1) в уравнение (1.8) получаем характеристическое уравнение
$$
(\eta-\mu)\Phi(\eta,\mu)=\dfrac{\eta}{\sqrt{\pi}z_0}
\int\limits_{-\infty}^{\infty}
\exp(-{\mu'}^2)\Phi(\eta,\mu')d\mu'.
\eqno{(2.2)}
$$
Если принять нормировку
$$
\int\limits_{-\infty}^{\infty}
\exp(-{\mu'}^2)\Phi(\eta,\mu')d\mu'\equiv z_0,
\eqno{(2.3)}
$$
то уравнение (2.2) имеет решение
$$
\Phi(\eta,\mu)=\dfrac{1}{\sqrt{\pi}}\eta P\dfrac{1}{\eta-\mu}+
e^{\eta^2}\lambda(\eta)\delta(\eta-\mu),
\eqno{(2.4)}
$$
где $-\infty<\eta, \mu <+\infty$.

Здесь $\delta(x)$ -- дельта--функция Дирака, символ $Px^{-1}$
означает главное значение интеграла при интегрировании $x^{-1}$,
$\lambda(z)$ -- дисперсионная функция, введенная равенством
$\lambda(z)=-i\omega_1+\lambda_0(z)$,
где
$$
\lambda_0(z)=\dfrac{1}{\sqrt{\pi}}\int\limits_{-\infty}^{\infty}
\dfrac{e^{-\tau^2}\tau d\tau}{\tau-z}.
$$

Собственные функции (2.4) называются собственными функциями непрерывного
спектра, ибо спектральный параметр $\eta$ непрерывным образом заполняет всю
действительную прямую.

Таким образом, собственные решения уравнения (2.3) имеют вид
$$
h_\eta(x_1,\mu)=\exp\Big(-\dfrac{x_1}{\eta}z_0\Big)
\Big[\dfrac{1}{\sqrt{\pi}}\eta P\dfrac{1}{\eta-\mu}+
\exp(\eta^2)\lambda(\eta)\delta(\eta-\mu)\Big].
\eqno{(2.5)}
$$

Собственные решения (2.5) отвечают непрерывному спектру характеристического
уравнения, ибо спектральный параметр непрерывным образом пробегает всю
числовую прямую.
По условию задачи мы ищем решение, невозрастающее вдали от стенки.
В связи с этим спектром граничной задачи будем называть положительную
действительную полуось параметра $\eta$.

Приведем формулы Сохоцкого для дисперсионной функции:
$$
\lambda^{\pm}(\mu)=-i\omega_1+\lambda_0(\mu)\pm is(\mu),
$$
где
$$
\lambda_0(\mu)=\dfrac{1}{\sqrt{\pi}}\int\limits_{0}^{\infty}
\dfrac{e^{-\tau^2}\tau d\tau}{\tau-\mu},\qquad s(\mu)=i\sqrt{\pi}\mu e^{-\mu^2}.
$$
Разложим дисперсионную функцию в ряд Лорана по отрицательным степеням
переменного $z$ в окрестности бесконечно удаленной точки:
$$
\lambda(z)=-i\omega_1-\dfrac{1}{2z^2}-\dfrac{3}{4z^4}-\dfrac{15}{8z^6}-\cdots,
\quad z\to \infty.
\eqno{(2.6)}
$$

Из разложения (2.6) видно, что при малых значениях $\omega_1$
дисперсионная функция имеет два отличающиеся лишь знаками комплексно--значных
нуля $\eta_0^\circ$ и $-\eta_0^\circ$, причем
$$
\eta_0^\circ(\omega_1)=\dfrac{1+i}{2\sqrt{\omega_1}}, \quad
\Re \eta_0^\circ >0.
\eqno{(2.7)}
$$

Отсюда видно, что при $\omega_1\to 0$ оба нуля дисперсионной функции
имеют пределом одну бесконечно удаленную точку $\eta_i=\infty$ кратности
(порядка) два.

Введем выделенную частоту колебаний пластины, ограничивающей газ:
$$
\omega_1^*=\max\limits_{0<\mu<+\infty}
\sqrt{-\lambda_0^2(\mu)+s^2(\mu)}\approx 0.733.
$$

Эту частоту колебаний будем называть {\it критической}. Введем индекс задачи
$\varkappa=\varkappa(G)=\frac{1}{2\pi}\arg G(t)\Big|_0^\infty$, где
$G(t)=\lambda^+(t)/\lambda^-(t)$.

В \cite{ALY-1} показано, что в случае, когда частота колебаний пластины
меньше критической, т.е. при $0\leqslant \omega <\omega_1^*$, индекс функции
$G(t)$ равен единице. Это означает \cite{ALY-1},
что число
комплексных нулей дисперсионной функции в плоскости с разрезом вдоль
действительной оси, равно двум.

В случае, когда частота колебаний пластины превышает критическую
($\omega>\omega_1^*$) индекс функции $G(t)$ равен нулю: $\varkappa(G)=0$.
Это означает, что дисперсионная функция не имеет нулей в верхней и нижней
полуплоскостях. В этом случае дискретных (частных) решений исходное
уравнение (1.8) не имеет.

Таким образом, дискретный спектр характеристического уравнения,
состоящий из нулей дисперсионной функции, в случае
$0\leqslant \omega_1<\omega_1^*$ есть множество из двух точек
$\eta_0$ и $-\eta_0$.
При $\omega_1>\omega_1^*$ дискретный спектр --- это пустое множество.
При $0\leqslant \omega_1<\omega_1^*$ собственными функциями
характеристического уравнения являются следующие два решения
характеристического уравнения:
$$
\Phi(\pm \eta_0,\mu)=\dfrac{1}{\sqrt{\pi}}\dfrac{\pm \eta_0}{\pm \eta_0-\mu}.
$$

Под $\eta_0$ будем понимать тот из нулей дисперсионной функции,
который обладает свойством:
$
\Re [{(1-i\omega_1)}/{\eta_0}]>0.
$
Для этого нуля убывающее собственное решение уравнения (1.8) имеет вид
$$
h_{\eta_0}(x_1,\mu)=
\exp\Big(-\dfrac{x_1z_0}{\eta_0}\Big)\Phi(\eta_0,\mu).
$$
Это означает, что дискретный спектр рассматриваемой граничной задачи
состоит из одной точки $\eta_0$ в случае $0 <\omega_1<\omega_1^*$.
При $\omega_1\to 0$ оба нуля, как уже указывалось выше, перемещаются в
одну и ту же бесконечно удаленную точку. Это значит, что в этом случае
дискретный спектр характеристического уравнения состоит из одной
бесконечно удаленной точки кратности два
и является присоединенным к непрерывному спектру. Этот спектр является
также и спектром рассматриваемой граничной задачи. Однако, в этом случае
дискретных (частных) решения ровно два:
$h_1(x_1,\mu)=1$ и $h_2(x_1,\mu)=x_1-\mu$.

\begin{center}
3. АНАЛИТИЧЕСКОЕ РЕШЕНИЕ ГРАНИЧНОЙ ЗАДАЧИ
\end{center}
Составим общее решение уравнения (1.8) в виде суммы частного
решения, убывающего вдали от стенки, и интеграла по
непрерывному спектру от собственных решений, отвечающих непрерывному спектру:
$$
h(x_1,\mu)=a_0\exp\Big(-\dfrac{x_1z_0}{\eta_0}\Big)\Phi(\eta_0,\mu)+
\int\limits_{0}^{\infty}
\exp\Big(-\dfrac{x_1z_0}{\eta}\Big)\Phi(\eta,\mu)a(\eta)d\eta.
\eqno{(3.1)}
$$
Здесь $a_0$ -- неизвестный постоянный коэффициент, называемый коэффициентом
дискретного спектра, причем при $\varkappa=0$ этот коэффициент равен нулю,
$a(\eta)$ -- неизвестная функция, называемая коэффициентом
непрерывного спектра, $\Phi(\eta,\mu)$ -- собственные функции
характеристического уравнения,
отвечающие непрерывному спектру и нормировке (2.3). Функция $a(\eta)$
подлежит нахождению из граничных условий (1.9)--(1.11).

Разложение (3.1) можно представить в явном виде:
$$
h(x_1,\mu)=\dfrac{\eta_0a_0}{\sqrt{\pi}(\eta_0-\mu)}\exp\Big(-\dfrac{x_1z_0}
{\eta_0}\Big)+
$$
$$
+\dfrac{1}{\sqrt{\pi}}\int\limits_{0}^{\infty}
\exp\Big(-\dfrac{x_1z_0}{\eta}\Big)\dfrac{\eta a(\eta)d\eta}{\eta-\mu}+
\exp\Big(-\dfrac{x_1z_0}{\mu}+\mu^2\Big)\lambda(\mu)a(\mu)\theta_+(\mu),
\eqno{(3.2)}
$$
где $\theta_+(\mu)$ -- функция Хэвисайда, $\theta_+(\mu)=1,\mu>0$,
$\theta_+(\mu)=0,\mu<0$.

Очевидно, что разложение (3.2)
автоматически удовлетворяет граничному условию (1.11) вдали
от стенки. Подставим разложение (3.2) в граничное условие (1.10).
Получаем одностороннее сингулярное интегральное уравнение с ядром Коши
$$
\dfrac{\eta_0a_0}{\sqrt{\pi}(\eta_0-\mu)}+\dfrac{1}{\sqrt{\pi}}
\int\limits_{0}^{\infty}
\dfrac{\eta a(\eta)d\eta}{\eta-\mu}+
\exp(\mu^2)\lambda(\mu)a(\mu)\theta_+(\mu)=2(U_0q+d),\; \mu>0.
\eqno{(3.3)}
$$

Введем вспомогательную функцию
$$
N(z)=\dfrac{1}{\sqrt{\pi}}\int\limits_{0}^{\infty}\dfrac{\eta a(\eta)d\eta}
{\eta-z}.
\eqno{(3.4)}
$$

Пользуясь формулами Сохоцкого для вспомогательной и дисперсионной функций,
от уравнения (3.3) приходим к краевому условию:
$$
\lambda^+(\mu)\bigg[N^+(\mu)-2(U_0q+d)+
\dfrac{\eta_0a_0}{\sqrt{\pi}(\eta_0-\mu)}\bigg]-
$$
$$
=\lambda^-(\mu)\bigg[N^-(\mu)-2(U_0q+d)+
\dfrac{\eta_0a_0}{\sqrt{\pi}(\eta_0-\mu)}\bigg]=0, \quad \mu>0.
\eqno{(3.5)}
$$

Рассмотрим соответствующую однородную краевую задачу Римана:
$$
X^+(\mu)=G(\mu)X^-(\mu),\quad
\qquad\mu>0.
\eqno{(3.6)}
$$

Решение задачи Римана было рассмотрено в \cite{ALY-2} и дается интегралом
типа Коши:
$$
X(z)=\dfrac{1}{z^\varkappa}\exp V(z),
\eqno{(3.7)}
$$
где $V(z)$ понимается как интеграл типа Коши
$$
V(z)=\dfrac{1}{\pi}\int\limits_{0}^{\infty}
\dfrac{[q(\tau)-\pi\varkappa]d\tau}{\tau-z}, \eqno{(3.8)}
$$
где
$$
q(\tau)=\dfrac{\ln G(\tau)}{2i}=\dfrac{\arg G(\tau)}{2}-
\dfrac{i}{2}\ln |G(\tau)|.
$$
Под $\ln G(\tau)$ понимается главное значение логарифма
$\ln G(\tau)=\ln|G(\tau)|+i\arg G(\tau)$, фиксированное в нуле условием
$\ln G(0)=0$. Заметим, что автоматически выполняется условие
$\ln G(+\infty)=i\arg G(+\infty)$.

Вернемся к решению неоднородной задачи (3.5), предварительно преобразовав
с помощью (3.6) ее к виду:
$$
X^+(\mu)\bigg[N^+(\mu)-2(U_0q+d)+\dfrac{\eta_0a_0}{\sqrt{\pi}(\eta_0-\mu)}\bigg]-
$$
$$
-X^-(\mu)\bigg[N^-(\mu)-2(U_0q+d)+
\dfrac{\eta_0a_0}{\sqrt{\pi}(\eta_0-\mu)}\bigg]=0,
\quad \mu>0.
\eqno{(3.9)}
$$

Учитывая поведение всех входящих в краевое условие (3.9)
функций в комплексной плоскости и в бесконечно удаленной точке получаем
с помощью равенств (3.7) и (3.8) общее решение, из которого находим
$$
N(z)=2(U_0q+d)+\dfrac{\eta_0a_0}{\sqrt{\pi}(z-\eta_0)}+
\dfrac{1}{X(z)}\Big[C_0+\dfrac{C_1}{z-\eta_0}\Big],
\eqno{(3.10)}
$$
где $C_0, C_1$ -- произвольные постоянные, причем при $\varkappa=0\; C_1=0$,
а при $\varkappa=1\; C_0=0$.

Пусть $\varkappa=1$. Их условия $N(\infty)=0$ находим, что $C_1=-2(U_0q+d)$.
Полюс в точке $\eta_0$ у решения (3.10) устраним условием
$a_0\eta_0/\sqrt{\pi}+C_1/X(\eta_0)=0$, из которого находим:
$$
a_0=\dfrac{2\sqrt{\pi}(U_0q+d)}{\eta_0 X(\eta_0)}.
\eqno{(3.11)}
$$

Для нахождения коэффициента непрерывного спектра воспользуемся формулой
Сохоцкого для функции (3.4):
$$
N^+(\mu)-N^-(\mu)=2\sqrt{\pi}i \mu a(\mu),\qquad \mu>0.
\eqno{(3.12)}
$$

Найдем граничные значения решения (3.10) сверху и снизу на действительной
полуоси и подставим их в равенство (3.12). Находим, что
$$
\sqrt{\pi}i \eta a(\eta)=i(U_0q+d)\Big[\dfrac{1}{X^+(\mu)}-
\dfrac{1}{X^-(\mu)}\Big]=2(U_0q+d)\dfrac{\sin\zeta(\eta)}{X(\eta)},
\eqno{(3.13)}
$$
где
$$
\zeta(\eta)=q(\eta)-\pi\varkappa, \qquad \varkappa=0,1.
$$
В случае $\varkappa=0$, как нетрудно видеть, коэффициент $a(\eta)$ вычисляется
также по формуле (3.13).

Коэффициенты дискретного и непрерывного спектров разложения (3.1) (или (3.2))
найдены и определяются равенствами (3.11) и (3.13).
На этом этапе доказательство разложения (3.1) (или (3.2)) закончено.

Остается найти постоянную $d$, входящую в граничное условие (1.5) (или (1.10)).
Нам понадобится первый момент функции
распределения, который находим с помощью характеристического уравнения:
$$
\int\limits_{-\infty}^{\infty}e^{-\mu^2}\mu \Phi(\eta,\mu)d\mu=-i\omega_1\eta.
$$
С помощью дисперсионного уравнения аналогично устанавливается равенство:
$$
\int\limits_{-\infty}^{\infty}e^{-\mu^2}\mu \Phi(\eta_0,\mu)d\mu=
-i\omega_1\eta_0.
$$
Подставим разложение (3.1) в граничное условие (1.10). Учитывая два предыдущих
равенства, получаем уравнение:
$$
i\omega_1\Big[a_0\eta_0+\int\limits_{0}^{\infty}\eta a(\eta)d\eta\Big]=
-q\Big(U_0-\dfrac{d}{1-q}\Big).
\eqno{(3.14)}
$$
Пусть $\varkappa=1$. Подставим (3.11) и (3.13) в уравнение (3.14).
Получаем уравнение:
$$
i\omega_12\sqrt{\pi}(U_0q+d)\Bigg[\dfrac{1}{X(\eta_0)}+
\int\limits_{0}^{\infty}
\dfrac{\sin \zeta(\eta)d\eta}{X(\eta)(\eta-\eta_0)}\Bigg]=
-q\Big(U_0-\dfrac{d}{1-q}\Big).
\eqno{(3.15)}
$$
Интеграл из уравнения (3.15) можно вычислить с помощью интегрального
представления из \cite{ALY-2}:
$$
\dfrac{1}{X(z)}=z-V_1-\dfrac{1}{\pi}\int\limits_{0}^{\infty}
\dfrac{\sin \zeta(\tau)d\tau}{X(\tau)(\tau-z)},
\eqno{(3.16)}
$$
где
$$
V_1=-\dfrac{1}{\pi}\int\limits_{0}^{\infty}\zeta(\tau)d\tau.
$$
С помощью интегрального представления (3.16) уравнение (3.15) превращается в
алгебраическое уравнение, из которого находим
$$
d=(1-q)U_0\dfrac{1-2\sqrt{\pi}i\omega_1(V_1-\eta_0)}{1+Q_1},
$$
где
$$
Q_1= 2\sqrt{\pi}\dfrac{1-q}{q}i\omega_1(V_1-\eta_0).
$$

Пусть теперь $\varkappa=0$. Рассуждая аналогично и используя
интегральное представление из \cite{ALY-2}:
$$
\dfrac{1}{X(z)}=1-\dfrac{1}{\pi}\int\limits_{0}^{\infty}
\dfrac{\sin \zeta(\eta)d\eta}{X(\eta)(\eta-z)},
\eqno{(3.17)}
$$
получаем, что
$$
d=(1-q)U_0\dfrac{1-2\sqrt{\pi}i\omega_1V_1}{1+Q_0},
$$
где
$$
Q_0=2\sqrt{\pi}\frac{1-q}{q}i\omega_1 V_1.
$$

На основании найденных выражений для $d$ получаем, что
$2U_0q+d={U_0}/({1+Q_\varkappa})$, где
$$
Q_\varkappa= 1+2\frac{1-q}{q}i\omega_1\sqrt{\pi}(V_1-\varkappa \pi),\qquad
\varkappa=0,1.
$$

\begin{center}
4. ФУНКЦИЯ РАСПРЕДЕЛЕНИЯ
\end{center}

Установленное разложение (3.2) означает, что функция распределения
газовых молекул построена. Рассмотрим функцию распределения летящих к стенке
молекул ($\mu<)$):
$$
h(x_1,\mu)=\dfrac{a_0\eta_0}{\sqrt{\pi}(\eta_0-\mu)}e^{-x_1z_0/\eta_0}+
\dfrac{1}{\sqrt{\pi}}\int\limits_{0}^{\infty}e^{-x_1z_0/\eta}
\dfrac{\eta a(\eta)d\eta}{\eta-\mu}.
\eqno{(4.1)}
$$ Рассмотрим случай $\varkappa=0$. В этом случае из (4.1) получаем:
$$
h(x_1,\mu)=\dfrac{1}{\sqrt{\pi}}\int\limits_{0}^{\infty}e^{-x_1z_0/\eta}
\dfrac{\eta a(\eta)d\eta}{\eta-\mu},
$$
или, в явном виде
$$
h(x_1,\mu)=\dfrac{2U_0}{1+Q_0}\dfrac{1}{\pi}\int\limits_{0}^{\infty}
e^{-x_1z_0/\eta}\dfrac{\sin \zeta(\eta)d\eta}{X(\eta)(\eta-\mu)}.
\eqno{(4.2)}
$$
Интеграл из (4.2) при $x_1=0$ вычислим аналитически с помощью интегрального
представления (3.17). В результате получаем значение функции распределения
летящих к стенке молекул непосредственно у стенки:
$$
h(0,\mu)=\dfrac{2U_0}{1+Q_0}\Big[1-\dfrac{1}{X(\mu)}\Big],\qquad \mu<0.
\eqno{(4.3)}
$$

В случае $\varkappa=1$ аналогично рассуждая, получаем:
$$
h(0,\mu)=\dfrac{2U_0}{1+Q_1}\Big[1-\dfrac{1}{(\mu-\eta_0)X(\mu)}\Big],\qquad \mu<0.
\eqno{(4.4)}
$$
Нуль $\eta_0$ дисперсионной функции из (4.4) может быть найден из формулы
\cite{ALY-2}
$$
\lambda(z)=i\omega_1(z^2-\eta_0^2)X(z)X(-z).
\eqno{(4.5)}
$$
Согласно (4.5) для вычисления $\eta_0$ получаем следующее выражение:
$$
\eta_0=\sqrt{z_*^2-\dfrac{\lambda(z_*)}{i\omega_1X(z_*)X(-z_*)}},
$$
причем точку $z_*$ для численных расчетов удобнее брать на мнимой оси.

\begin{center}
5. СКОРОСТЬ ГАЗА В ПОЛУПРОСТРАНСТВЕ И НЕПОСРЕДСТВЕННО У СТЕНКИ
\end{center}

Безразмерная скорость газа в полупространстве вычисляется по формуле:
$$
U_y(t_1,x_1)=\dfrac{1}{\pi^{3/2}}\int e^{-C^2}C_y
[1+C_yh(x_1,C_x)e^{-i\omega_1t_1}]d^3C,
$$
или, через функцию $h(x_1,\mu)$:
$$
U_y(t_1,x_1)=\dfrac{e^{-i \omega_1t_1}}{2\sqrt{\pi}}\int\limits_{-\infty}^{\infty}
e^{-\mu'^2}h(x_1,\mu')d\mu'.
\eqno{(5.1)}
$$
Подставим в (5.1) разложение (3.1). Воспользовавшись нормировкой (2.3), получаем:
$$
U_y(t_1,x_1)=\dfrac{e^{-i\omega_1t_1}z_0}{2\sqrt{\pi}}
\Bigg[a_0e^{-x_1z_0/\eta_0}+\int\limits_{0}^{\infty}
e^{-x_1z_0/\eta}a(\eta)d\eta\Bigg].
\eqno{(5.2)}
$$
Подставим в (5.2) коэффициенты дискретного и непрерывного спектров. Начнем
со случая $\varkappa=1$. Тогда безразмерная скорость газа в полупространстве
равна:
$$
U_y(t_1,x_1)=\dfrac{U_0e^{-i\omega_1t_1}z_0}{1+Q_\varkappa}
\Bigg[\dfrac{e^{-x_1z_0/\eta_0}}{\eta_0 X(\eta_0)}+\dfrac{1}{\pi}
\int\limits_{0}^{\infty}e^{-x_1z_0/\eta}\dfrac{\sin\zeta(\eta)d\eta}
{\eta X(\eta)(\eta-\eta_0)}\Bigg].
\eqno{(5.3)}
$$
Интеграл из (5.3) можно вычислить аналитически при $x_1=0$.
С помощью представления (3.16) получаем, что
$$
U_y(t_1,x_1)=\dfrac{U_0e^{-i\omega_1t_1}z_0}{1+Q_\varkappa}\Bigg(1+
\dfrac{1}{\eta_0 X(0)}\Bigg).
\eqno{(5.4)}
$$

Из формулы для факторизации дисперсионной функции \cite{ALY-2} находим, что
$X(0)=\sqrt{i z_0/\omega_1\eta_0^2}$. Следовательно, для размерной скорости
газа из (5.4) находим:
$$
u_y(t_1,0)=u_0|A_1|e^{-i(\omega_1t_1-\varphi_1)},
\eqno{(5.5)}
$$
где
$$
A_1=\dfrac{z_0+\sqrt{\omega_1z_0}e^{-i\pi/4}}{1+Q_1},\qquad
\varphi_1=\arg A_1.
$$

Пусть теперь индекс задачи равен нулю.
Подставим (3.1) в (5.2) и поменяем порядок интегрирования. Затем,
используя нормировочное соотношение (2.3), приходим к равенству:
$$
U_y(x_1,t_1)=\dfrac{e^{-i\omega_1t_1}z_0}{2\sqrt{\pi}}
\int\limits_{0}^{\infty}\exp\Big(-\dfrac{x_1}{\eta}z_0\Big)a(\eta)d\eta.
$$
Теперь воспользуемся формулой (3.13) для коэффициента непрерывного спектра.
Получаем что массовая скорость газа в полупространстве равна:
$$
U_y(x_1,t_1)=U_0e^{-i\omega_1t_1}\dfrac{z_0}{1+Q_0}\dfrac{1}{\pi}
\int\limits_{0}^{\infty}\exp\Big(-\dfrac{x_1}{\eta}z_0\Big)
\dfrac{\sin\zeta(\eta)}{\eta X(\eta)}d\eta.
\eqno{(5.6)}
$$

Вычислим значение массовой скорости непосредственно вблизи у стенки.
Для вычисление интеграла из (5.6) при $x_1=0$ воспользуемся интегральным
представлением из \cite{ALY-2}. Получаем, что массовая скорость в
полупространстве вычисляется по формуле:
$$
U_y(0,t_1)=U_0e^{-i\omega_1t_1}\dfrac{z_0}{1+Q_0}\Big(1-\dfrac{1}{X(0)}\Big).
\eqno{(5.7)}
$$

Для нахождения величины факторизующей функции в нуле воспользуемся теперь
формулой факторизации дисперсионной функции \cite{ALY-2}:
$
\lambda(z)=\lambda_\infty X(z)X(-z),
$
где
$
\lambda_\infty=\lambda(\infty)=-i\omega_1.
$
Замечая, что $\lambda(0)=1-i\omega_1$, находим:
$
X^2(0)={\lambda(0)}/{\lambda_\infty}=1+i \nu/\omega.
$
Согласно (5.7) значение скорости газа у стенки равно:
$$
U_y(0,t_1)=U_0e^{-i\omega_1t_1}A_0,\quad\text{где}\quad A_0=(1-i\omega_1)
\dfrac{\sqrt{\omega_1+i}-\sqrt{\omega_1}}{\sqrt{\omega_1+i}(1+Q_0)}.
$$
Значение размерной скорости непосредственно у стенки дается выражением:
$$
u_y(0,t)=u_0|A_0|e^{-i\omega_1 t_1-\varphi_0},\qquad \varphi_0=\arg A_0.
\eqno{(5.8)}
$$
Формулы (5.5) и (5.8) можно объединить в одну:
$$
u_y(0,t)=u_0|A_\varkappa|e^{-i\omega_1 t_1-\varphi_\varkappa},\qquad
\varkappa=0,1.
$$

\begin{center}
6. О ГИДРОДИНАМИЧЕСКОМ ХАРАКТЕРЕ РЕШЕНИЯ
\end{center}

Покажем, что при малых $\omega_1$ решение (5.3) переходит в решение,
приведенное в \cite{LandauG}:
$$
v=u_0e^{-x/\delta}e^{i(x/\delta-\omega t)}.
\eqno{(6.1)}
$$

Здесь
$
\delta=\sqrt{{2\nu_k}/{\omega}},
$
$\nu_k$ -- кинематическая вязкость газа. Формула (6.1) выведена для случая
сплошной среды в случае, когда ограничивающая среду плоскость совершает
гармонические колебания по закону $u_s(t)=u_0e^{-i\omega t}$.

Воспользуемся формулой (5.3).
При малых $\omega_1$ нуль дисперсионной функции $\eta_0\to \infty$,
следовательно, интеграл по непрерывному спектру является исчезающе малым.
Далее заметим, что
$
\eta_0X(\eta_0)=e^{V(\eta_0)},
$
а при больших значений $|\eta_0|$ интеграл $V(\eta_0)$ исчезает при малых
$\omega_1$. Значит, при $\omega_1\to 0$ для скорости газа получаем выражение
$$
u_y(x_1,t_1)=u_0e^{-i\omega_1t_1}e^{-x_1/\eta_0}.
\eqno{(6.2)}
$$

Здесь $\omega_1t_1=\omega t, x_1=\nu \sqrt{\beta}x$.
При малых $\omega_1$ для нуля дисперсионной функции имеем:
$
\eta_0=(1+i)/(2\sqrt{\omega_1}).
$
Поэтому выражение (6.2) преобразуется следующим образом:
$
u_y(x,t)=u_0e^{-i\omega t}e^{-x\sqrt{\beta}/\tau\eta_0}.
$
Замечая, что для используемой БГК--модели кинетического уравнения
${\tau}/{\beta}=2\nu_k$, далее получаем:
$$
\dfrac{\nu\sqrt{\beta}}{\eta_0}=
\dfrac{\sqrt{\beta}2\sqrt{\omega\tau}}{\tau(1+i)}=
\dfrac{1-i}{\sqrt{\tau/(\omega \beta)}}=
\dfrac{1-i}{\sqrt{(2\nu_k)/\omega}}=\dfrac{1-i}{\delta}.
$$
Это означает, что
$
u_y(x,t)=u_0e^{-i\omega t}e^{-(1-i)x/\delta},
$
что в точности совпадает с выражением (24,5) из \cite{LandauG}.

\begin{center}
7. СИЛА ТРЕНИЯ, ДЕЙСТВУЮЩАЯ НА КОЛЕБЛЮЩУЮСЯ ГРАНИЦУ
\end{center}
Компонента тензора вязких напряжений, приходящаяся на единицу площади
колеблющейся границы, вычисляется по формуле
$$
\sigma_{xy}(t)=m\int v_xv_yf(t,0,\mathbf{v})d^3v.
\eqno{(7.1)}
$$
Согласно \cite{LandauG} сила трения (приходящаяся на единицу площади),
действующая со стороны газа на пластину, равна
$
F_s(t)=-\sigma_{xy}(0,t).
$
Поэтому согласно (7.1)
$$
F_s(t_1)=
-e^{-i \omega_1t_1}\dfrac{p}{\sqrt{\pi}}
\int\limits_{-\infty}^{\infty}e^{-\mu^2}\mu h(0,\mu)d\mu, \quad p=nkT.
\eqno{(7.2)}
$$

Подставим в (7.2) разложение (3.1). Получаем, что
$$
F_s(t_1)=-i\omega_1e^{-i\omega_1t_1}\dfrac{p}{\sqrt{\pi}}\Big[a_0\eta_0+
\int\limits_{0}^{\infty}\eta a(\eta)d\eta\Big].
\eqno{(7.3)}
$$
Рассмотрим случай, когда индекс задачи равен единице. Подставим в (7.3)
коэффициенты дискретного и непрерывного спектров:
$$
F_s(t_1)=\dfrac{2U_0pi\omega_1}{1+Q_1}e^{-i\omega_1t_1}\Big[\dfrac{1}{X(\eta_0)}
+\dfrac{1}{\pi}\int\limits_{0}^{\infty}\dfrac{\sin \zeta(\eta)d\eta}
{X(\eta)(\eta-\eta_0)}\Big].
\eqno{(7.4)}
$$
Теперь воспользуемся интегральным представлением (3.16).
Тогда соотношение (7.4) преобразуется к виду:
$$
F_s(t_1)=\dfrac{2U_0pe^{-i\omega_1t_1}i\omega_1}{1+Q_1}(V_1-\eta_0).
\eqno{(7.5)}
$$
Пусть теперь индекс задачи равени нулю. Тогда из формулы (7.3) получаем:
$$
 F_s(t_1)=\dfrac{2U_0pi\omega_1e^{-i\omega_1t_1}}{1+Q_1} \dfrac{1}{\pi}
 \int\limits_{0}^{\infty}\dfrac{\sin\zeta(\eta)d\eta}{X(\eta)}.
$$
Пользуясь представлением (3.17), отсюда получаем:
$$
F_s(t_1)=- \dfrac{2U_0pi\omega_1e^{-i\omega_1t_1}}{1+Q_1}V_1.
\eqno{(7.6)}
$$
Формулы (7.5) и (7.6) можно объединить в одну:
$$
F_s(t_1)=-F_\varkappa e^{-i(\omega_1t_1-\varphi_\varkappa)},\qquad \varkappa=0,1.
\eqno{(7.7)}
$$
В (7.7) $F_s$ -- модуль силы трения, $\varphi_\varkappa$ -- сдвиг ее фазы,
$$
F_s=\dfrac{|\omega_1(V_1-\eta_0\varkappa)|}{|1+O_\varkappa|}, \quad
\varphi_\varkappa=\arg(V_1-\eta_0\varkappa)-\dfrac{\pi}{2}-
\arg(1+Q_\varkappa),\quad \varkappa=0,1.
$$

\begin{center}
8. СИЛА ТРЕНИЯ КАК ФУНКЦИЯ ЧИСЛА КНУДСЕНА
\end{center}
Введем число Кнудсена как отношение $\Kn=l/\delta$, где $l$ -- средняя
длина свободного пробега газовых молекул, $\delta$
введена в п. 6. Выбирая длину свободного пробега согласно Черчиньяни
\cite{Cerc73}
как $l=\sqrt{\pi \beta}\eta/\rho$, $\eta$ -- динамическая вязкость газа,
получаем, что
$$
\Kn=\dfrac{\sqrt{\pi}\sqrt{\beta}v_k \sqrt{\omega}}{\sqrt{2\nu_k}}=
\sqrt{\dfrac{\pi}{2}\omega}\sqrt{\nu_k\beta}.
$$
Замечая, что $2\nu_k={1}/({\nu \beta})={\tau}/{\beta}$, находим, что
$$
\Kn=\dfrac{\sqrt{\pi}}{2}\sqrt{\omega_1}.
$$
Найдем
выражение силы трения в гидродинамическом пределе, т.е. при $\omega\to 0$.
В этом пределе нуль дисперсионной функции переходит в выражение
$\eta_0^\circ=\dfrac{1+i}{2\sqrt{\omega_1}}$,
а коэффициент $V_1\to V_1^\circ$, где $V_1^\circ$ -- величина, пропорциональная
коэффициенту изотермического скольжения из задачи Крамерса об изотермическом
скольжении,
$$
V_1^\circ=-\dfrac{1}{\pi}\int\limits_{0}^{\infty}
\Big[\arcctg\dfrac{\lambda_0(\tau)}{s(\tau)}
-\pi\Big]d\tau=1.016.
$$
Для силы трения в гидродинамическом пределе (опуская длительные выкладки)
получаем:
$$
F_s(t_1)=-2\sqrt{\dfrac{2}{\pi}}U_0p e^{-i(\omega_1t_1+\pi/4)} \times $$$$\times
\dfrac{\Kn-(1-i)\dfrac{2}{\sqrt{\pi}}{\Kn\!}^2V_1^\circ}
{1+2\dfrac{1-q}{q}\sqrt{2}e^{-i\pi/4}
\big[\Kn-(1-i)\dfrac{2}{\sqrt{\pi}}{\Kn\!}^2V_1^\circ\big]},
$$
или, в виде отрезка ряда по степеням числа Кнудсена,
$$
F_s(t_1)=-2\sqrt{\dfrac{2}{\pi}}U_0p e^{-i(\omega_1t_1+\pi/4)}
\Big[\Kn-(1-i)\Big(\dfrac{2}{\sqrt{\pi}}V_1^\circ+
2\dfrac{1-q}{q}\Big){\Kn\!}^2\Big] \eqno{(8.1)}
$$

Выясним связь формулы (8.1) с предыдущим результатом \cite{DudkoD}, где
рассмотрен гидродинамический режим со скольжением. Для этого
введем величину $L=C_m\Kn$, где $C_m=\dfrac{2}{\sqrt{\pi}}V_1^\circ$ --
коэффициент изотермического скольжения, найденный Черчиньчни \cite{Cerc73},
и рассмотрим случай $L\ll 1$. В этом случае формула (8.1) преобразуется к виду:
$$
F_s(t_1)=-U_0p\sqrt{2\omega \tau}
\Big[1-(1-i)\Big(L+\dfrac{1-q}{q}\sqrt{\pi\omega\tau}\Big)\Big]
e^{-i(\omega_1t_1+\pi/4)}.
\eqno{(8.2)}
$$
Заметим, что если обозначить
$$
L_1=\dfrac{1-q}{q}\sqrt{\pi\omega\tau},
$$
то
$$
1-(1-i)(L+L_1)=\sqrt{1-2(L+L_1)+2(L+L_1)^2}
\exp\Big(i\arctg\dfrac{L+L_1}{1-L-L_1}\Big).
$$
При условии, что $L\ll 1$, т. е. $L\ll 1$ и $L_1\ll 1$, отсюда получаем, что
$1+(1-i)(L+L_1)=(1-L-L_1)e^{i(L+L_1)}$. Следовательно, согласно (8.2) получаем:
$$
F_s(t_1)=-U_0p\sqrt{2\omega\tau}(1-L-L_1)e^{-i(\omega_1t_1+\pi/4-L-L_1)}.
\eqno{(8.3)}
$$
Нетрудно заметить, что эта формула в точности совпадает с формулой
$$
F_s(t_1)=\sqrt{2}u_0\dfrac{\nu_k\rho}{\delta}(1-L)e^{-i(\omega_1t_1+\pi/4-L)}
$$
из диссертации \cite{DudkoD}, если положить в (8.3) $q=1$, т. е. $L_1=0$.

\begin{center}
9. СВОБОДНО МОЛЕКУЛЯРНЫЙ РЕЖИМ
\end{center}

Рассмотрим случай, когда индекс задачи равен нулю и $\omega_1\gg 1$,
или, что эквивалентно условию $\Kn\gg 1$, т. е. свободно молекулярный режим.
Возьмем выражение (7.7) для силы трения. Рассмотрим выражение для
коэффициента $V_1$:
$$
V_1=-\dfrac{1}{2\pi i}\int\limits_{0}^{\infty}\ln G(\tau)d\tau.
$$
Умножим это равенство на $i\omega_1$ и представим его в явном виде:
$$
i\omega_1V_1=-\dfrac{1}{2\pi}\int\limits_{0}^{\infty}\omega_1
\ln\dfrac{-i\omega_1+\lambda_0(\tau)+is(\tau)}
{-i\omega_1+\lambda_0(\tau)-is(\tau)}d\tau.
$$
Разложим по степеням $\omega_1^{-1}$ подынтегральную функцию и ограничимся
первыми тремя членами:
$$
\omega_1\ln\dfrac{-i\omega_1+\lambda_0(\tau)+is(\tau)}
{-i\omega_1+\lambda_0(\tau)-is(\tau)}=-2s(\tau)+\dfrac{2i}{\omega_1}s(\tau)
\lambda_0(\tau)+\dfrac{2}{\omega_1^2}[s(\tau)\lambda_0^2(\tau)-
\dfrac{1}{3}s^3(\tau)].
$$
Отсюда получаем, что
$$
i\omega_1V_1= 0.282-\dfrac{i}{\omega_1}0.053-\dfrac{1}{\omega_1^2}0.022.
$$
Подставим это разложение в соотношение (7.7),
получаем, что сила трения в этом режиме равна:
$$
F_s(t_1)=-2pU_0e^{-i\omega_1t_1}q\dfrac{0.282-{i}{\omega_1^{-1}}0.053-
{\omega_1^{-2}}0.022}{q+2\sqrt{\pi}(1-q)[0.282-{i}{\omega_1^{-1}}0.053-
{\omega_1^{-2}}0.022]}.
$$

\begin{center}
10. МОЩНОСТЬ ДИССИПАЦИИ ЭНЕРГИИ
 \end{center}
Рассмотрим вопрос о диссипации энергии колеблющейся пластины.
Рассмотрим мощность диссипации энергии, т.е. величину диссипации
энергии в единицу времени, приходящуюся на единицу площади колеблющейся
пластины.

Формулу (7.7) для силы трения представим в виде
$F_s(t_1)=-e^{-i\omega_1t_1}\Phi_\varkappa$,
где
$$
\Phi_\varkappa=\dfrac{2U_0p}{1+Q_\varkappa}i\omega_1(V_1-\eta_0\varkappa),\qquad
\varkappa=0,1.
$$
 Согласно \cite{LandauE} усредненная по времени мощность
диссипации энергии вычисляется по формуле
$$
W=\dfrac{1}{2}\Re\Big(u_0\Phi_\varkappa^*\Big)=\dfrac{U_0}{2\sqrt{\beta}}
\Re \Phi_\varkappa^*.
\eqno{(10.1)}
$$

В формуле (10.1) звездочка ($*$) означает комплексное сопряжение.

Формулу (10.1) представим в явном виде:
$$
W=W_0\Re\Bigg[\dfrac{i\omega_1(V_1-\eta_0\varkappa)}{1+2\sqrt{\pi}
\dfrac{1-q}{q}i\omega_1(V_1-\eta_0\varkappa)}\Bigg]^*, \eqno{(10.2)}
$$
где
$$
W_0=\dfrac{U_0^2p}{\sqrt{\beta}}, \qquad \varkappa=0,1.
$$

Проведем краткий графический анализ полученных результатов.
Кривые $1,2,3$ на фигурах 1--4
отвечают значениям коэффициента аккомодации $q=1,0.75,0.5$, а на фигуре 5
-- значениям $q=1,0.5,0.25$.

Зависимость величины амплитуды и сдвига фазы скорости газа от частоты
колебаний ограничивающей газ плоскости непосредственно у стенки изображена
соответственно на фиг. 1 и 2.

Зависимость величины амплитуды силы трения от частоты колебаний ограничивающей
газ плоскости изображена на фиг. 3.

Зависимость величины сдвига фазы силы трения от частоты ограничивающей газ
плоскости изображена на фиг. 4.

Зависимость мощности диссипации энергии от частоты колебаний
ограничивающей газ пластины согласно (10.2) изображена на фиг. 5.

\begin{center}
 ЗАКЛЮЧЕНИЕ
\end{center}
В настоящей работе сформулирована и решена аналитически вторая задача
Стокса --- задача о поведении
разреженного газа, занимающего полупространство над стенкой, совершающей
гармонические колебания. Рассматриваются аккомодационные граничные условия
Черчиньяни.
Используется уравнение, полученное в результате
линеаризации модельного кинетического уравнения Больцмана.
На основе аналитического решения построены функция распределения и
скорость разреженного газа в
полупространстве и непосредственно у стенки. Выяснен гидродинамический
характер скорости газа. Отыскивается также сила
трения, действующая со стороны газа на пластину.
Сила трения представлена в виде отрезка ряда по числу
Кнудсена, а также исследована в свободно молекулярном режиме.
Наконец, найдена мощность диссипации
энергии, приходящаяся на единицу площади колеблющейся границы.

В дальнейшем авторы намерены решить аналитически вторую задачу Стокса
с использованием кинетического уравнения Шахова \cite{Shakhov} или
эллипсоидально--статистического уравнения \cite{Cerc73}, приводящих
к правильному числу Прандтля.

\begin{figure}[h]
\begin{center}
\includegraphics[width=16cm, height=10cm]{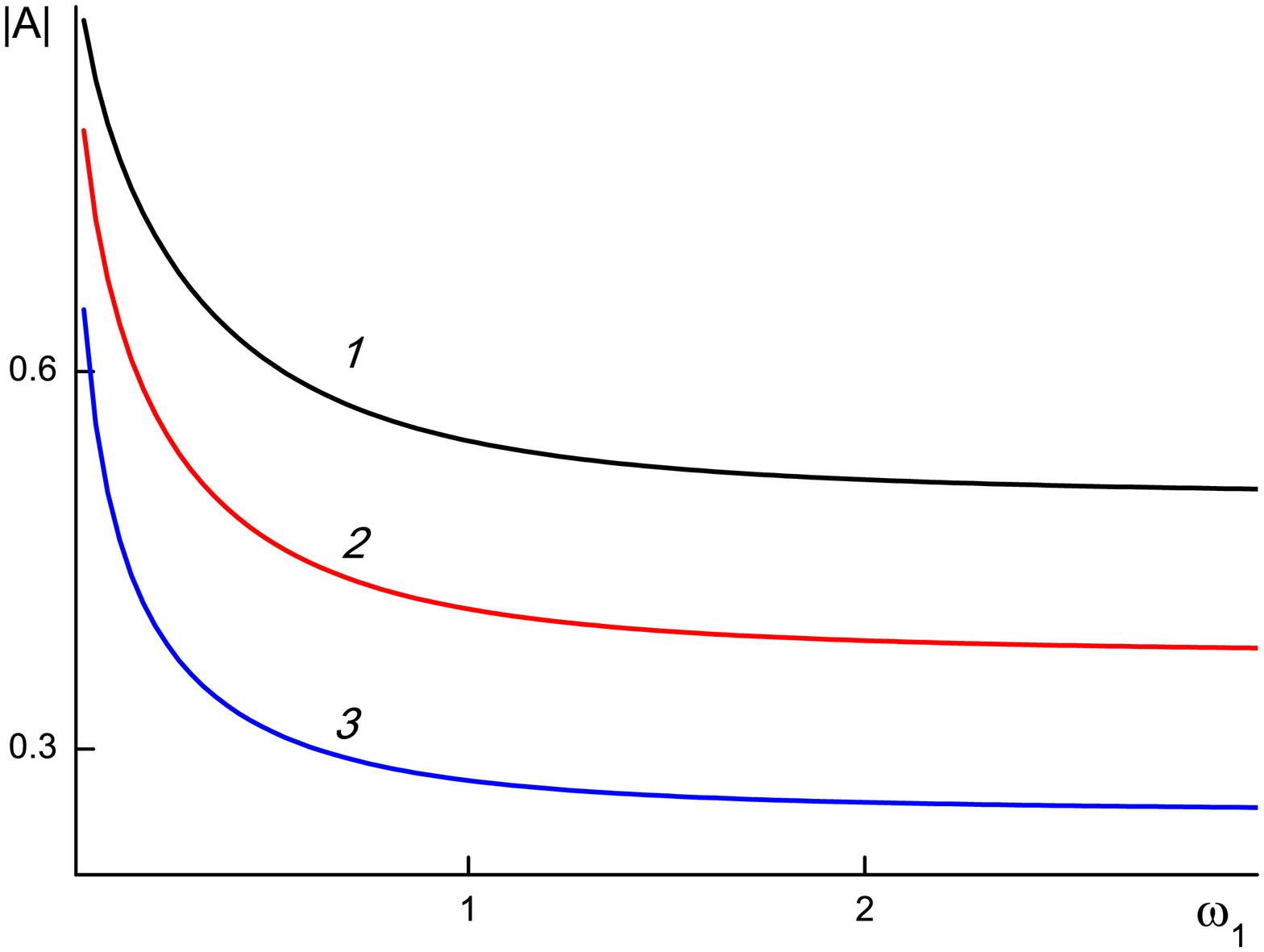}
{Фиг. 1.}
\end{center}
\end{figure}
\begin{figure}[h]
\begin{center}
\includegraphics[width=16cm, height=10cm]{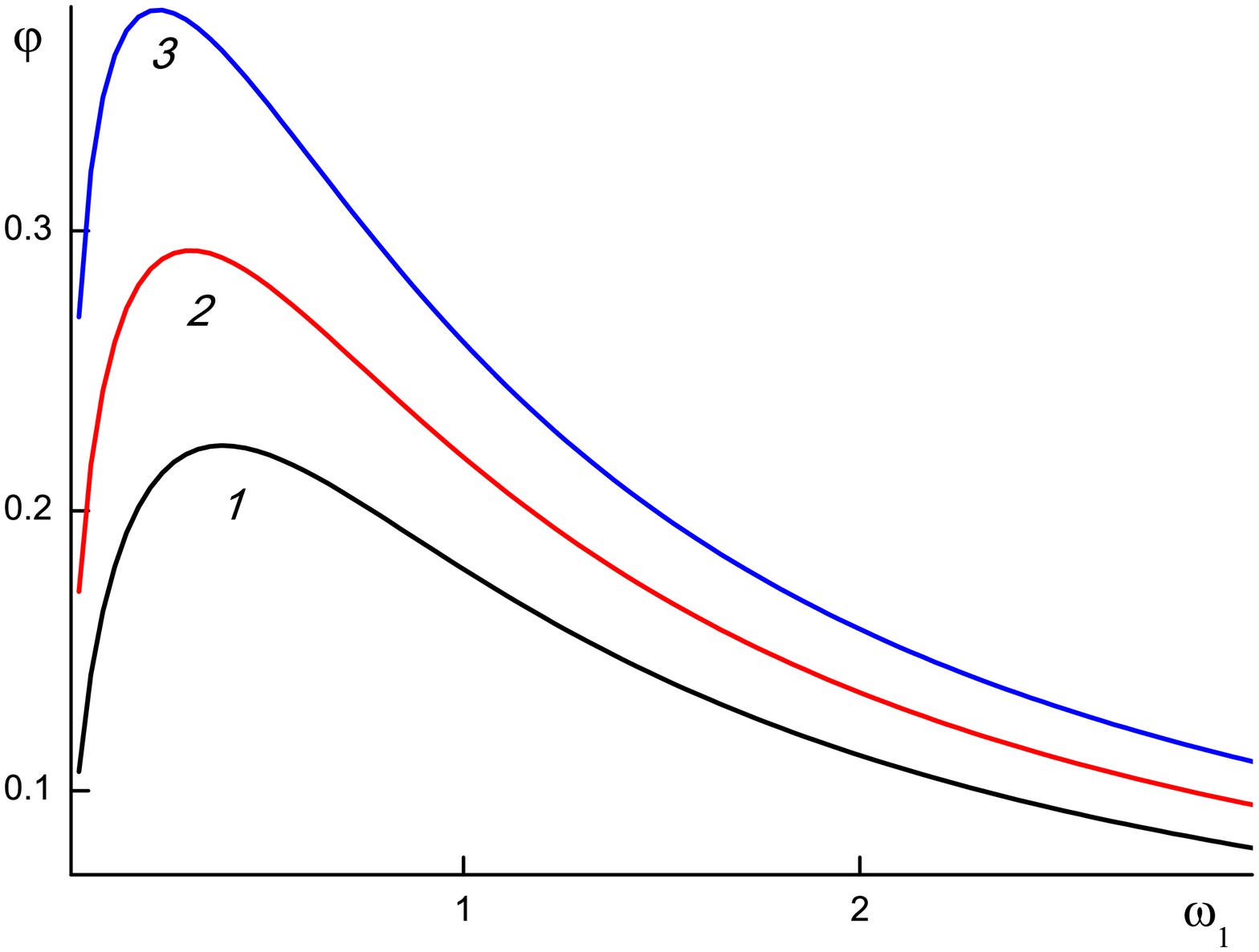}
{Фиг. 2.}
\end{center}
\end{figure}

\begin{figure}[t]
\begin{center}
\includegraphics[width=16cm, height=10cm]{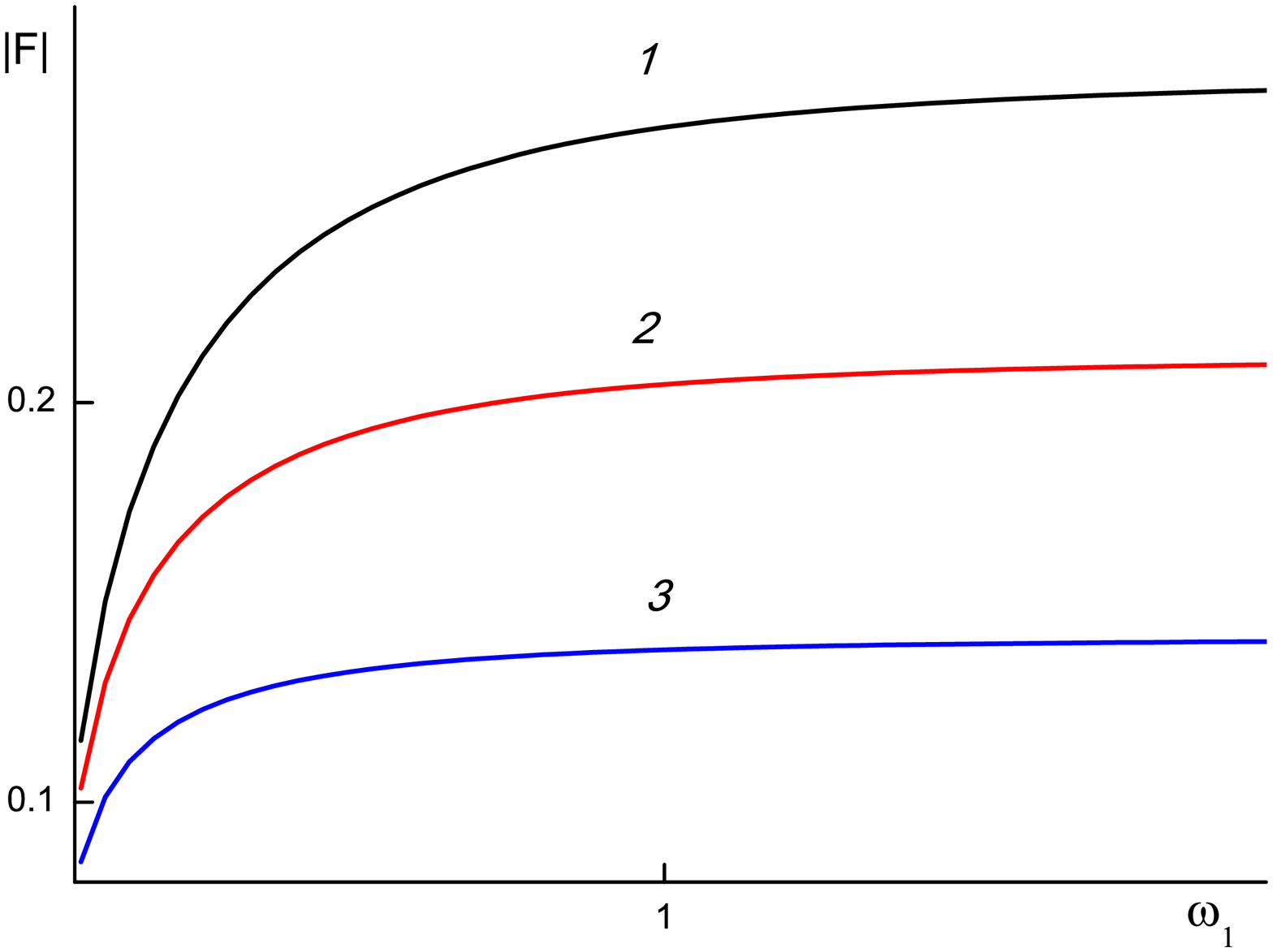}
{Фиг. 3.}
\end{center}
\end{figure}

\begin{figure}[h]
\begin{center}
\includegraphics[width=16cm, height=10cm]{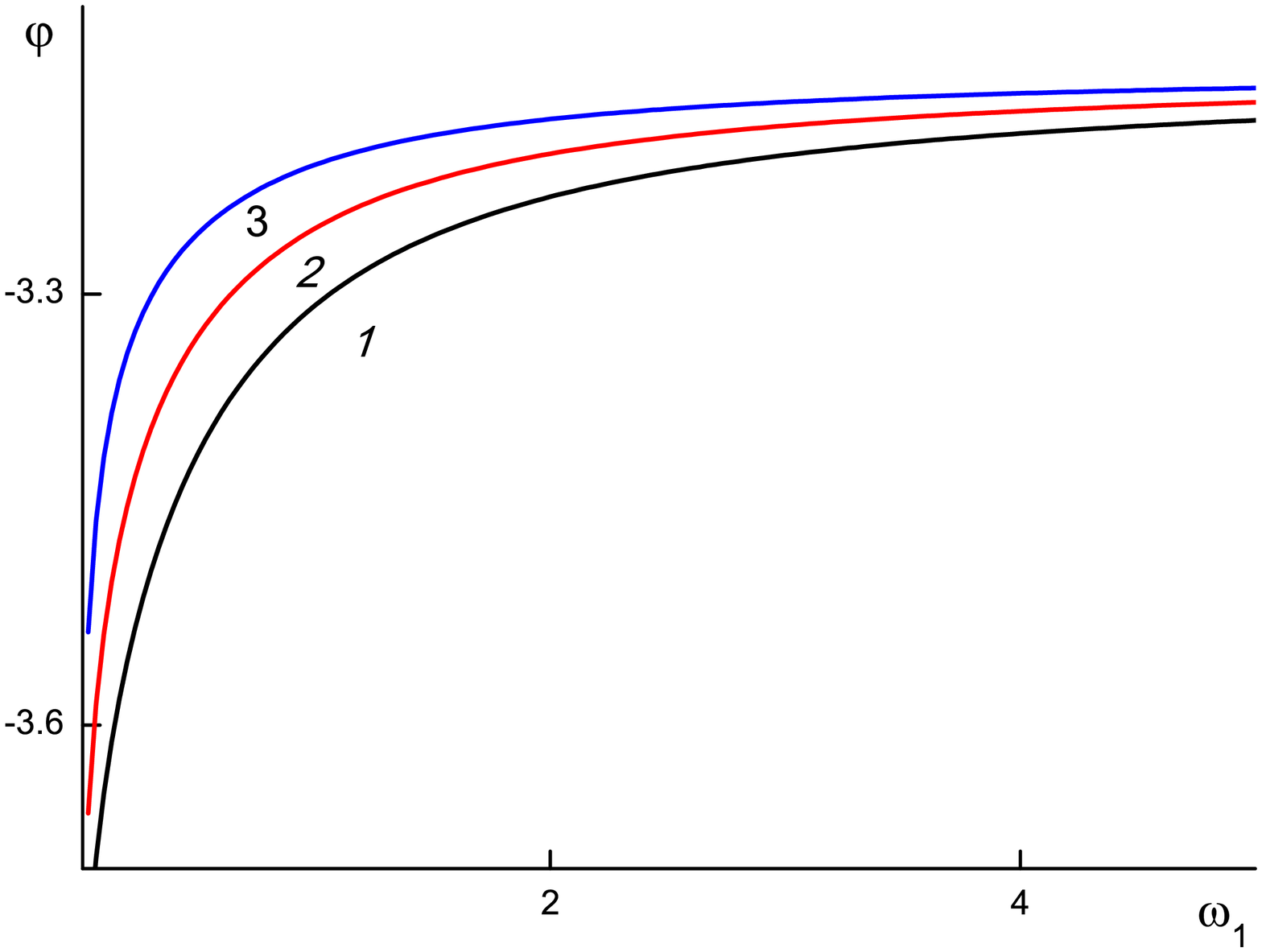}
{Фиг. 4.}
\end{center}
\end{figure}

\begin{figure}[h]
\begin{center}
\includegraphics[width=16cm, height=10cm]{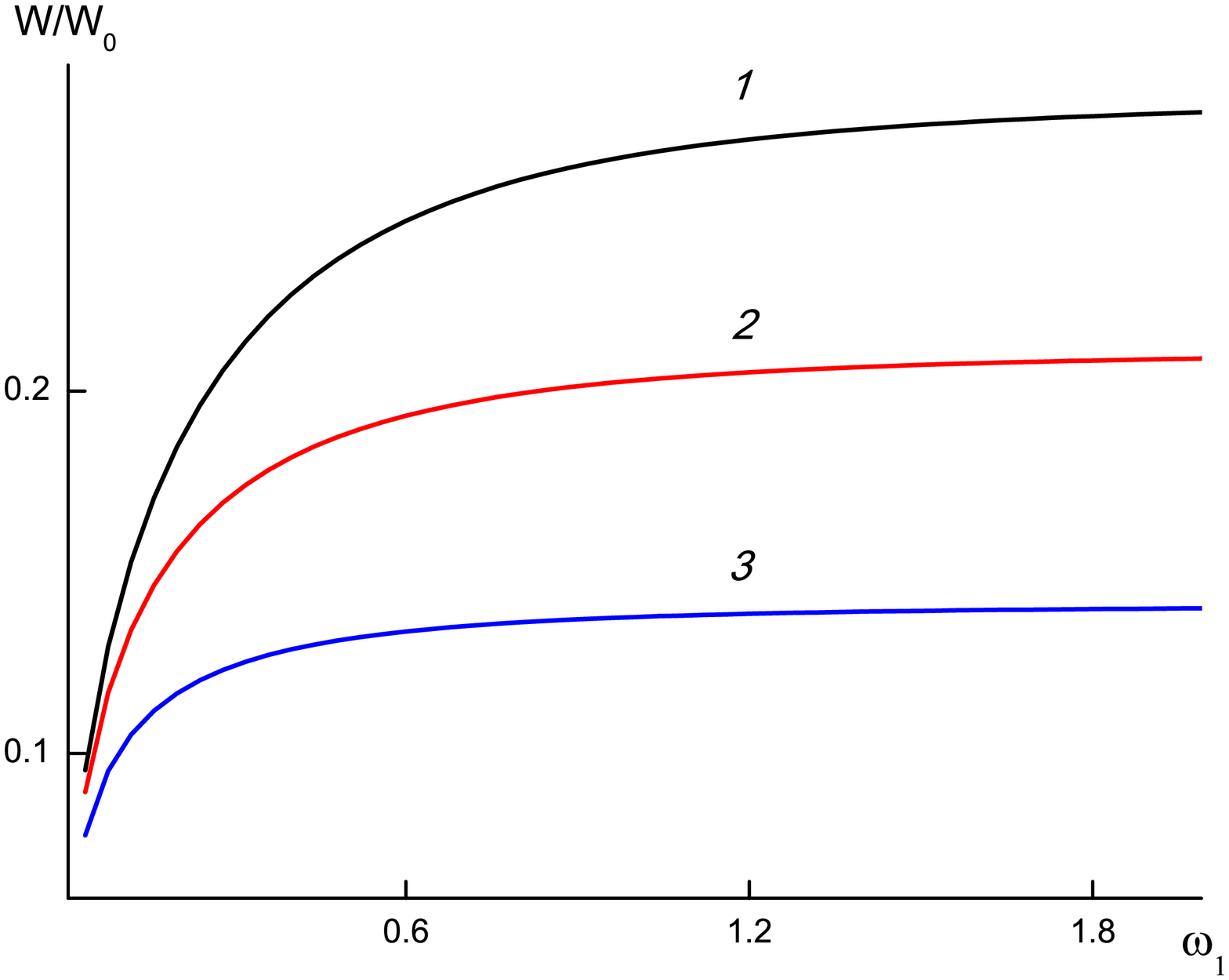}
{Фиг. 5.}
\end{center}
\end{figure}

\end{document}